\documentclass[reqno]{amsart}
\usepackage{amssymb,latexsym,amscd}
\theoremstyle{plain}

\newtheorem{proposition}{Proposition}
\newtheorem{corollary}{Corollary}
\theoremstyle{definition}
\newtheorem{definition}{Definition}
\newtheorem{remark}{Remark}
\newtheorem{example}{Example}

\numberwithin{equation}{section}

\newcommand{\beq }{\begin{equation}}
\newcommand{\eeq }{\end{equation}}

\begin{document}
\begin{abstract}
In this work we investigate the material point model
(MP-model) and exploit the geometrical meaning of the "entropy form"
introduced by B.Coleman and R.Owen (\cite{CO}). We analyze full and partial integrability (closeness) condition of the entropy form for the 
model of thermoelastic point and for the the deformable ferroelectric crystal media point. We show that the extended thermodynamical space
introduced by R.Hermann and widely exploited by R. Mrugala with his collaborators and other researchers, extended possibly by time, with its
 canonical contact structure is an appropriate setting for the development of material point models in different physical situations. 
This allows us to formulate the model of a material point and the corresponding entropy form in terms similar to those of the homogeneous
 thermodynamics, \cite{Mru}.  Closeness condition of the entropy
 form is reformulated as the requirement that the admissible processes curves belongs to the constitutive surface $\Sigma$  of the model.
 Our principal result is the description of the constitutive surfaces of the material point model as the Legendre submanifolds $\Sigma_{S}$
 (equilibrium submanifolds of homogeneous thermodynamics) of the space $\mathcal{P}$ \emph{shifted by the flow of Reeb vector
 field}. This shift is controlled, at the points of Legendre submanifold $\Sigma_{S}$ by the \emph{entropy production function} $\sigma$.
\end{abstract}
\title[Geometry of the entropy form.]%
{Material point model and the geometry of the entropy form.}
\author{M.Dolfin}\address{Department of Mathematics, University of Messina, Messina, It.}
\email{dolfin@dipmat.unime.it}
\author{S.Preston}
\address{Department of Mathematics and Statistics, Portland State
University, Portland, OR, USA} \email{serge@mth.pdx.edu}
\author{L.Restuccia}
\address{Department of Mathematics, University of Messina, Messina, It.}
\email{lrest@dipmat.unime.it} \maketitle

\today
 \tableofcontents

\section{Introduction.}

The goal of this work is to investigate the material point model
(MP-model) and exploit the geometrical meaning of the "entropy form"
introduced by B.Coleman and R.Owen (\cite{CO}) and, later on, applied to different physical systems in \cite{CDFR,DFR1,DFR2, DFRe, FRR}. 

In their work, B.Coleman and R.Owen developed the basic dynamical scheme
of thermodynamical systems postulating the basic properties of
states and processes in a thermodynamical systems, introduced the abstract notion of action, Clausius-Duhem
 inequality etc.\par
  Geometrical structure of their model was later on reformulated in terms of the bundle theory in \cite{DFR1,DFR2,DFRe}. \par
The second part of their work is devoted to illustrations and
applications of their scheme to the theory of "simple material
elements", including the "elastic points".  The authors introduced
the \emph{"entropy form"}, defined in the appropriate \emph{state
space} (extended by time variable) and determining the change of entropy
produced by the flux along the process defined by the curve in the
state space. Entropy form is constructed starting from the balance of entropy 
of continuum thermodynamics.  Applying the
energy balance and some kinematical relations for the rate of change
of the involved dynamical variables, one rewrites the entropy
increase due to the flux as the integral of a 1-form $\eta'$ in the
state space. \emph{Entropy function} is then defined as an
\emph{upper potential of the entropy form}.  Such a potential
exists, in particular, in the case where entropy 1-form is closed.
Closure conditions provides some set of constitutional relations for
participating fields. The study of the entropy form and of the
corresponding constitutive relations given by the condition of
closeness of the entropy form for different thermodynamical systems
were done in the works \cite{DFR1,DFR2,DFRe,FRR} and some
others.\par
In this work we would like to analyze the geometrical meaning of the
entropy form and that of its integrability.\par

In the first part we revisit the "basic model" of the thermoelastic material
point of Coleman-Owen in order to determine which part of the
constitutive relations for the dynamical system of MP-model can be
obtained from different integrability conditions of the entropy form
and what information should be added from the continuum
thermodynamics in order to construct the closed dynamical system of
MP-model. We also reformulate the "material point-entropy form"
model both for the "basic model" of thermoelastic point and for the deformable ferroelectric crystal media \cite{FRR}, in
terms of \emph{extended thermodynamical phase space}. We study
integrability conditions of the entropy form in the way similar to
the study of Coleman-Owen simple model. We determine the
constitutive part of the integrability conditions and the dynamical
part, entering the dynamical system of the model.
\par

 It is easy to observe strong similarities of the model of the material point here
 with the geometrical formalism of homogeneous thermodynamics (\cite{H, Mru}, etc.)  Exploiting these similarities,
 in the second part of this work we identify the phase space of a material point with the extended (by time) version of the
 thermodynamical phase space $\mathcal{P}$ of Caratheodory-Herman-Mrugala and the entropy form -  with the dynamical part of the contact form in this space.
 This allows us to formulate the geometrical theory of material point and the corresponding entropy form
 in terms similar to those of the homogeneous thermodynamics, \cite{Mru}.  In particular, closeness condition of the entropy
 form is reformulated as the requirement that the admissible processes curves belongs to the constitutive surface $\Sigma$  of the model.
 Our principal result is the description of the constitutive surfaces of the material point model as the Legendre submanifolds $\Sigma_{S}$
 (equilibrium submanifolds of homogeneous thermodynamics) of the space $\mathcal{P}$ \emph{shifted by the flow of Reeb vector
 field}. This shift is controlled, at the points of Legendre submanifold $\Sigma_{S}$ by the \emph{entropy production function} $\sigma$. \par

Ideally, the construction of a material point model (MP-model) in specific physical circumstances
should start by specifying the basic state space of physical fields
and their spacial derivatives (gradients, divergences) whose time
evolution one would like to study using the MP-model   (notice that
in that respect the MP-model is similar to the "Extended
Thermodynamics", \cite{MR}).  Then one would like to determine the
\emph{dynamical system for these variables} and, if necessary , to
complement it with the constitutive relations closing the system.
All this should be done in such a way that the energy balance law
and the II law of thermodynamics expressed in the Clausius-Duhem
inequality would be satisfied in a natural (for MP-model) form. In
the Conclusion we mention some possible directions of this
development, leaving its realization to the future work.

\vskip1cm
 \addtocontents{toc}{\textbf{Part I. Coleman-Owen model, entropy form and integrability.}}
\centerline{\textbf{Part I. Coleman-Owen model of thermoelastic point, entropy form and
integrability.}}

In this part we define and analyze the entropy form in the model of
material point (MPM) suggested by B.Coleman and R.Owen, \cite{CO} and
later on studied in numerous works \cite{DFR1,DFR2}, etc. We will
present our analysis on the example of thermoelastic material point
used by B.Coleman and R.Owen as the basic model system.

\section{Entropy form of a thermoelastic system.}

In this section we remind the construction of the \textbf{entropy
form} introduced in (\cite{CO}) and studied in \cite{DFR2}. We start
with a balance of entropy of a continuum thermodynamical system in
the form

\beq\label{entrobal} \dot s +\frac{1}{\rho}\nabla\cdot
\mathbf{J}_{S}=\Xi, \eeq

where $s$ is specific entropy density, $\mathbf J_{S}$ is the
entropy flux and $\Xi$ is the entropy production that, due to the
II law of thermodynamics, is nonnegative.  Thus, we assume that
the entropy supply is zero (the system is adiabatically isolated
).\par We also admit the relation between the entropy flux
$\mathbf J_{S}$ and the heat flux $ \mathbf q$ in the form

\beq \mathbf{J}_{S}=\theta^{-1}\mathbf{q}+\mathbf{k}, \eeq where
$\mathbf{k}$ is the extra entropy flux that will be taken to be
zero in this section but will appear in more complex situations
below.\par

Plugging in the last expression, we rewrite (\ref{entrobal}) in the
form

\beq\label{entrobal1} \dot s +\frac{1}{\rho}\nabla\cdot
(\theta^{-1}\mathbf{q})=\dot s +\frac{1}{\theta \rho}\nabla\cdot
\mathbf{q} + \frac{1}{ \rho}\mathbf{q}\cdot \nabla\theta^{-1}=\Xi.
\eeq

\emph{Elastic} material body is considered as a 3-dim material
manifold $M^3$ embedded at the time $t$ into the physical
(euclidian) space $E^3$ by the diffeomorphism $\phi_{ t}:
M\rightarrow E^3$.  Deformation of $M$ is characterized by the
deformation gradient $F^{i}_{I}=\frac{\partial \phi_t^i}{\partial
X^I}$).

We admit the internal energy balance for a thermoelastic material
point in the form \beq \rho  \dot{\epsilon }=p_{(i)}-\nabla \cdot
\mathbf{q}=\boldsymbol\sigma :\mathbf D-\nabla \cdot \mathbf{q}.
\eeq where $\epsilon$ is the internal energy per unit of volume,
 $\mathbf D=(\nabla \mathbf{v})^{s}$ is the symmetrized strain rate tensor
($\mathbf v$ being the velocity at the material point $m$),
$\sigma^{ij}$ is the Cauchy stress tensor and $p_{(i)}$ is the
work power of the stress. Here and below the symbol : is used for
the contraction of tensors. \par
Expressing $\nabla \cdot
\mathbf{q}$ from the energy balance and noticing also that
$\mathbf D=(\mathbf F^{-1}\dot{\mathbf F})^s$ is equal to the
symmetrical part of the velocity gradient tensor $\mathbf
L=\mathbf F^{-1}\dot{\mathbf F}$ we finally present the entropy
balance in the form

\[
\dot s -\frac{1}{\theta } \dot{\epsilon}+\frac{1}{\theta
\rho}(\boldsymbol\sigma:\mathbf F^{-1}:\dot{\mathbf F})+\frac{1}{
\rho}\mathbf{q}\cdot \nabla\theta^{-1}=\Xi.
\]
Here we have used the angular momentum balance equation in the form
$\boldsymbol\sigma^T=\boldsymbol\sigma$.\par

Now, we refer this relation to a material element, i.e. small
enough volume of a material to associate with it definite values
of the state variables participating in the energy and the entropy
balance equations with the configuration, stress etc. see
\cite{CO}.\par

For such an element the Coleman-Owen model suggests a state space
in a way that presumably guarantees the completeness of the
dynamical system for the variables in its state space (not yet
chosen!) and describing for a given physical situation the
response of the material element to the exterior influence.\par

In such a case the evolution of the material properties is described
by the collection of scalar, vectorial and tensor functions of time

\[(\rho(t),\mathbf F(t),\theta(t),\mathbf{q}(t),\epsilon(t),
\boldsymbol{\sigma} (t), (\nabla \theta^{-1})(t), etc.),\]
 forming the
\emph{process} $\chi$ at the chosen material point. Variables listed
above are related by some constitutive relations determined by the
properties of the corresponding material media. Their time evolution
has to be determined by a dynamical system. It is imperative for the
closeness of the model to determine all such relations and to use
them for the reduction of the dynamical system to as simple one as
possible.  It is also important for the construction of dynamical
models of the material point, see, for instance \cite{Ha,Ha2}.
\par

The infinitesimal \emph{entropy production} along a process $\chi$
is given by the integral of 1-form

\begin{multline}
ds-\frac{1}{\theta } \dot{\epsilon}dt+\frac{1}{\theta
\rho}(\boldsymbol\sigma:\mathbf F^{-1}:\dot{\mathbf F})dt+\frac{1}{
\rho} \mathbf{q}\cdot \nabla\theta^{-1}dt=\\ =ds-\frac{1}{\theta }
d\epsilon+\frac{1}{\theta \rho}\boldsymbol\sigma:\mathbf
F^{-1}:d\mathbf F+\frac{1}{ \rho}\mathbf{q}\cdot
\nabla\theta^{-1}dt.
\end{multline}

Second expression in (2.6) can be considered as the exterior 1-form
(\emph{entropy action form}, see \cite{DFR2,FRR})

\beq \eta'=\frac{1}{\theta } d\epsilon-\frac{1}{\theta
\rho}\boldsymbol\sigma:\mathbf F^{-1}:d\mathbf F-\frac{1}{
\rho}\mathbf{q}\cdot \nabla\theta^{-1}dt \eeq
 for the process
$\gamma(t)$ considered as a time parameterized curve in the
appropriate state space, i.e. the space of variables $(\epsilon,
\mathbf F, t)$ and other variables either independent in the proper
state space or determined by the constitutive relations.

 The \textbf{entropy action} at the material point \emph{along a process} $\gamma(\tau )$ from the time $\tau=0$ to $\tau=t$ is defined (postulated) as the integral

 \beq
 \Delta s (\gamma_t)=\int_{\gamma_t}\eta' =\int_{0}^{t}\gamma^{*}\eta'.
 \eeq

 Here $\gamma^{*}\eta'$ is the pullback of 1-form $\eta'$ to the interval of time.  As defined, $\Delta s (\gamma_t)$ is the functional of the curve
 $\gamma(\tau)$ with $\tau\in (0,t)$.
 So far nothing guarantees that the \emph{entropy action} is defined \emph{in the space of processes} $\gamma$ connecting two points in the state space.
 In particular it is unclear when the \emph{entropy function} is defined as the function in the space of
 variables associating with a material point, up to  an arbitrary constant (due to a choice of initial point).
 Assuming that the state space is simply connected this question rely on the property of the form $\eta$ to be closed, in fact if
 $d\eta =0$ then locally (and in a simply connected space, globally)there exists the potential $U$
 \beq
 \eta=dU,\ \Delta U(\gamma)=U(\gamma(end))-U(\gamma(start)),
 \eeq
 i.e. the function $U$ of state variables (including time) defined up to an arbitrary constant.
 Notice that the potential $U$
 is defined by the entropy flux only and as a result coincides with the entropy (up to a constant)
 only when the  entropy production in an admissible processes
 is zero.  \par
 If the potential $U$ and the entropy function $s$ both exist as the functions in the space where processes $\gamma$ are studied, and if at the initial moment
 we normalize the potential $U$ by the condition $U(\gamma (0))=s(\gamma (0))$, the difference

 \beq\label{prod}
 \sigma=s(\gamma (t))-U(\gamma(t))
 \eeq
 \emph{is equal to the entropy production $\sigma$ during the process} $\gamma (\tau), \tau \in[0,t]$.

\begin{remark} Coleman and Owen postulated fulfillment of the II law of Thermodynamics in
the inequality form that is weaker then the equality (\ref{prod})

\beq S(\gamma(end))-S(\gamma(start))\geqq
\int_{0}^{t}\gamma^{*}\eta'. \eeq

This property of upper semi-continuity of the entropy function $S$
(see\cite{CO}, Sec.10) leads to the restriction on the space of
direction of admissible processes. Namely, inequality (2.10) can be
rewritten in the form

\beq\label{prod1} \int_{0}^{t}\gamma^{*}(dS-\eta') \geqq 0. \eeq

If one has equality in this relation, both $\gamma(t)$ and the
inverse process $\gamma(-t)$ are thermodynamically admissible.
But if the inequality in (\ref{prod1}) is strict, then the inverse
process is prohibited. Since any segment of a thermodynamically
admissible process has to be admissible, this conclusion can be
localized, i.e. at each point $m$ in the state space $M$, there is
a cone $C_{m}\subset T_{m}(M)$ of thermodynamically admissible
directions. So, the II law requires the existence of a field of
tangent cones in the state space defining directions of admissible
processes. In continuum thermodynamics this is remedied by the
"Amendment to the Second Law", see \cite{ME,Mus}. In the material
point model this leads to some interesting geometrical
consequences (see below).
\end{remark}

 That is why in the papers \cite{DFR1, DFR2, DFRe, FRR}
 the conditions for closeness of this form in different situations were studied rather then more
 abstract notion of upper potential that would require much deeper thermodynamical analysis.\par

\subsection{State space and thermodynamical phase space.}

 As the next step, we would like to note resemblance of the form $\omega =ds-\eta'$ with the standard Gibbs
 form of the homogeneous thermodynamics (see \cite{Ca} or the next section).  Only the last term in (2.7) is
 qualitatively different form the terms in the contact Gibbs 1-form.  To remedy this difference we suggest to use the heat vector field
 $\boldsymbol{\mathcal H}$ introduced by M.Biot (Bi).  The vector field ${\boldsymbol{\mathcal H}}$ is defined uniquely, up to a constant in time vector field,
  by the condition

 \beq
 \dot {\boldsymbol{\mathcal H}}= \mathbf{q}.
 \eeq
 Using this vector field in the expression (2.7) for the 1-form $\eta'$,  we transform it to the 1-form (non-degenerate by variables
 $\epsilon, \mathbf F,\boldsymbol{\mathcal H}$)

 \beq\label{oneform1}
 \eta =\frac{1}{\theta } d\epsilon-\frac{1}{\theta \rho}\boldsymbol\sigma:\mathbf F^{-1}:d\mathbf F-\frac{1}{ \rho} (\nabla\theta^{-1}):d\boldsymbol{\mathcal H}.
 \eeq

 At that point we introduce the configurational space of a material point
 \beq
 B=\{\epsilon, \mathbf F, \boldsymbol{\mathcal H}  \}\eeq

 and the thermodynamical phase space (TPS) of the variables
 \beq
 \mathcal{P}=\{s;q^1=\epsilon, q^2=\mathbf F, q^3=\boldsymbol{\mathcal H}; p_1=\theta^{-1},p_2=\frac{1}{\theta \rho}\boldsymbol\sigma:\mathbf F^{-1},p_3=
 \frac{1}{ \rho} (\nabla\theta^{-1}). \}
 \eeq

The exterior 1-form

\beq \omega =ds-\eta =ds-(\frac{1}{\theta }
d\epsilon-\frac{1}{\theta \rho}\boldsymbol\sigma:\mathbf
F^{-1}:d\mathbf F-\frac{1}{ \rho}
(\nabla\theta^{-1}):d\boldsymbol{\mathcal H}) \eeq

 is typically contact (due
to the functional independence of the variables ($q^i ,p_j$) and,
therefore, defines the \emph{contact structure} in the space
$\mathcal{P}$. \par

\section{Integrability (closeness) conditions.}

Here we invoke the closeness condition of the entropy form both in
$(\epsilon,\mathbf F,t)$ and $(\epsilon, \mathbf F, \mathcal {H})$
variables. In the Part II below it is shown that the closeness
conditions of the forms $\eta ,\eta'$ are special cases of the
integrability conditions for the contact structure in the extended
thermodynamical phase space.\par

For the beginning  we revisit these conditions of integrability
for the form $\eta'$, obtained in \ref{DFR2} and solve them.  We
remind that the basic phase space is $\{\epsilon,\mathbf
F,\boldsymbol\beta =-\frac{1}{\rho}\nabla (\theta^{-1}) \}$ and
the form $\eta'$ is
\[
\eta'=-\frac{\boldsymbol\sigma:\mathbf F^{-1}}{\theta}\cdot
d\mathbf F+\theta^{-1}d\epsilon +(\mathbf q\cdot
\boldsymbol\beta)dt.
\]

Closeness conditions of $\eta '$ are

\beq\label{closeness}
\begin{cases}
\partial_{F}(\theta^{-1})=\partial_{\epsilon}(\frac{\boldsymbol\sigma:\mathbf F^{-1}}{\theta}),
\\ \partial_{\beta}(\theta^{-1})=0, \\  \partial_{\beta}(\frac{\boldsymbol\sigma:\mathbf F^{-1}}{\theta})=0;\\
\partial_{\beta}(q\cdot \beta)=0,\\ \partial_{t}(\frac{\boldsymbol\sigma:\mathbf F^{-1}}{\theta})=-\partial_{F}(\mathbf q\cdot \boldsymbol\beta),
\\ \partial_{t}(\theta^{-1})=\partial_{\epsilon}(\mathbf q\cdot \boldsymbol\beta).
\end{cases}
\eeq

The first three equations have the general local solution of the form
\[
\begin{cases}
\theta^{-1}=\partial_\epsilon U+c_{1}(t),\\
-\frac{\boldsymbol\sigma:\mathbf F^{-1}}{\theta}=\partial_F
U+c_{2}(t),
\end{cases}
\]
for a function $U(\epsilon,\mathbf F,t)$ and arbitrary functions
$c_{1}(t),c_{2}(t)$.\par

Using these expression in the second triplet of equations one sees
that these three equations in (\ref{closeness})
 are equivalent to the following
presentation of the function $(\mathbf q\cdot \boldsymbol\beta):$

\beq (\mathbf q\cdot \boldsymbol\beta)=\frac{\partial U}{\partial
t}+c'_{2}(t)\cdot \mathbf F +c'_{1}\epsilon+c_{3}(t), \eeq with an
arbitrary function $c_{3}(t)$.\par

The function $U(\epsilon,\mathbf F,t)$ is the time-dependent
entropy form potential defined in the space of basic variables.
These formulas determine the following constitutive relations

\beq
\begin{cases}
\theta^{-1}=\partial_\epsilon U+c_{1}(t),\\
\boldsymbol\sigma:\mathbf F^{-1}=-\frac{\partial_F U+c_{2}(t)}{\partial_\epsilon U+c_{1}(t))},\\
(\mathbf q\cdot \boldsymbol\beta)=\partial_t
U+c'_{2}(t)\cdot\mathbf F +c'_{1}\epsilon+c_{3}(t).
\end{cases}
\eeq
\vskip 0.4cm

Repeating the same arguments for the representation $\eta$ of the
entropy form in the space of variables $\{\epsilon, \mathbf F,
\mathcal{H } \}$ we get the following

\begin{proposition}
For the model presented above with the space of basic fields
$\{\epsilon, \mathbf F, \mathcal{H } \}$ condition of integrability
of the entropy form $\eta$ given by (\ref{oneform1}) is equivalent
to the existence of a potential $U(\epsilon, \mathbf
F,\boldsymbol{\mathcal H})$ such that the following constitutive
relations hold

\beq
\begin{cases}
\theta^{-1}=\partial_{\epsilon}U,\\
\boldsymbol\sigma:\mathbf F^{-1}=-\rho (\partial_{\epsilon}U)\partial_{F}U,\\
\nabla(\theta^{-1})=-\rho \partial_{\boldsymbol{\mathcal H}}U.
\end{cases}
\eeq
\end{proposition}

\section{Partial integrability, dynamical equations and constitutive relations.}
\par

Here we consider the dynamical system for the basic variables
$\epsilon, \mathbf F, \boldsymbol{\mathcal H}$ reformulating the
dynamical system introduced by Coleman and Owen:

\beq\label{firstsystem}
\begin{cases}
\dot {\mathbf F} =\mathbf L \mathbf F,\\
\dot \epsilon =\rho^{-1}\boldsymbol\sigma:\mathbf D-\rho^{-1}\nabla \cdot \mathbf{q},\\
\dot {\boldsymbol\beta} =\boldsymbol\gamma.
\end{cases}
\eeq

Here $\boldsymbol\gamma$ is just the notation for the derivative
of $\boldsymbol\beta =-\rho^{-1}\nabla \theta^{-1}$.\par

As the first step we replace third equation by the relation defining
$\boldsymbol{\mathcal H}$:

\beq\label{secondsystem}
\begin{cases}
\dot {\mathbf F} =\mathbf L \mathbf F,\\
\dot \epsilon =\rho^{-1}\boldsymbol\sigma:\mathbf L-\rho^{-1}\nabla \cdot \mathbf{q},\\
 \mathcal{\dot H} =\mathbf{q}.
\end{cases}
\eeq

Before considering the full system of integrability conditions
(\ref{closeness}), let us look at the meaning of partial
integrability of the entropy form; first for individual terms
(which, of course, is always possible) and then for couples of
these terms.\par

More specifically, assuming the \emph{Fourier relation} between the heat flux and the temperature gradient we have the relations
\[
\begin{cases}
\mathcal{\dot{H}}=  \mathbf{q};\\
\nabla\theta^{-1}=k  \mathbf{q}.
\end{cases}
\]
Thus, if $\nabla\theta^{-1}$ can be determined as a function of
basic variables , one of the equations of the dynamical system
will be closed. Looking at the form $\eta$, see (\ref{oneform1})
(or at the equation (\ref{closeness})${}_{3}$ we notice the
\emph{integrability of the third term}, i.e. the possibility to
write
\[
-\frac{1}{ \rho} (\nabla\theta^{-1}):d\boldsymbol{\mathcal
H}=\partial_{\mathcal H}\xi d\boldsymbol{\mathcal H}
\]

with a function $\xi (\epsilon, \mathbf F, \boldsymbol{\mathcal H})$
that guarantees fulfillment of
$(\nabla\theta^{-1})=\rho\partial_{\mathcal H }\xi,$ and allows to
rewrite third equation in the system (4.2) in the closed form

\beq \mathcal{\dot{H}}=k^{-1}\rho\frac{\partial \xi}{\partial
\boldsymbol{\mathcal H}}(\epsilon, \mathbf F, \boldsymbol{\mathcal
H}). \eeq

 Thus, \textbf{integrability of the third term in (\ref{oneform1}) allows
to close the third dynamical equation of the system (4.2) in the
space of basic variables.}\par

\emph{Integrability of the first term} of the form $\eta$ is equivalent to the statement that

\[\label{firstterm}
\theta^{-1}=\partial_\epsilon W
\]
for some differentiable function $W(\epsilon, \mathbf
F,\boldsymbol{\mathcal H})$.

Reversing this relation we get the constitutive relation in the form
$\epsilon =\epsilon (\theta, \mathbf F, \boldsymbol{\mathcal H})$.

Looking at the systems (\ref{firstsystem}) and (\ref{secondsystem})
we see that the \textbf{integrability of the first term in
(\ref{oneform1}) delivers the basic constitutive relation}
presenting internal energy $\epsilon$ as the function of basic
variables $\mathbf F,\boldsymbol{\mathcal H}$ and temperature
$\theta$.\par

\emph{Integrability of the second term} of the form $\eta$ is
equivalent to statement that for some function $Z(\epsilon, \mathbf
F,\boldsymbol{\mathcal H})$,

\[
\frac{1}{\theta \rho}\boldsymbol\sigma=\partial_F Z:\mathbf F.
\]

It expresses the Cauchy stress as the function of basic variables.
\textbf{As a result, integrability of the second term in
(\ref{oneform1}) allows to determine the elastic part of the
evolution of the internal energy $\epsilon$ in the space of basic
variables.}

If we allow the form $\eta$ to be integrable in the first two terms,
functions $W,Z$ should coincide up to an arbitrary additive term,
i.e. $\lambda(F)$: $Z=W+\lambda(F)$ and, using the expression
(\ref{firstterm}) for $\theta^{-1}$, we get

\[
\frac{1}{\rho}\boldsymbol\sigma:\mathbf F^{-1}=(\partial_\epsilon
W)^{-1}\partial_F W.
\]
We will get similar simplifications assuming integration other
couples of terms in the expression (\ref{oneform1}) for the form
$\eta$.  Finally, if we assume the full integrability of $\eta$, we
get the system (4.2), in terms of a potential $U(\mathbf F, \epsilon
, \boldsymbol{\mathcal H})$ of the entropy form

\beq
\begin{cases}
\dot {\mathbf F}=\mathbf L\mathbf F,\\
\dot \epsilon =-\frac{\partial U}{\partial \epsilon}\frac{\partial U}{\partial F}:\mathbf L-\rho^{-1}\nabla \cdot \mathbf{q},\\
 \mathcal{\dot H} =k^{-1}\rho\frac{\partial U}{\partial \boldsymbol{\mathcal H}}.
\end{cases}
\eeq

Additionally we get the constitutive definition of temperature
through the potential $U$ as:

\beq \theta^{-1}=\frac{\partial U}{\partial \epsilon}. \eeq

Canonical equation for $\mathbf F$ -
\[
\dot {\mathbf F}=\mathbf L\mathbf F
\]
has pure kinematical meaning.\par

Thus, to close the dynamical system the for variables $(\epsilon,
\mathbf F,\boldsymbol{\mathcal H})$ one has to determine, by some
constitutive relations or phenomenologically, the fields $\mathbf
L=\nabla \mathbf{v}$ and $\nabla \cdot \mathbf{q}$.  Another way
would be to include them in the list of dynamical fields and
construct additional dynamical equations for them in the spirit of
rational extended thermodynamics, see \cite{MR}. The first way led
Noll to his definition of generalized processes\cite{TN}. The
arguments presented above show that the constitutive relations
required in the definition of Noll reduces to the heat propagation
constitutive relation (Fourier, Cattaneo, etc.)\par \vskip 1cm

\section{Entropy form in deformable ferroelectric crystal media}

Let us now recall the phenomenological model of an elastic
deformable ferroelectric crystal medium worked out in
\cite{Mau1}-\cite{Mau3},\cite{MP} in a suitable Galilean quasi-static
approximation. We assume that the medium is formed by $n$ molecular
species, each one of them giving rise to a field of electric
dipoles.  The total polarization per unit of mass is given by:

\begin{equation}
\boldsymbol{\pi}=\frac { \textbf{P}_{pol}}{\rho},
\end{equation}

where $\mathbf P_{pol}$ is the {\it total polarization} (per unit
of volume)

 The vector $\boldsymbol{\pi}$ is assumed to satisfy the
following  balance equation
\begin{equation}\label{pi}
I\ddot{\boldsymbol{\pi}}=\boldsymbol{\mathcal E}+^{\,\, L}\mathbf
E+\rho^{-1}(\nabla\cdot^{ \,\,L} \mathbb E),
\end{equation}

where $I\neq 0$ is an "inertia constant" (which in the following
will not be restrictive to let be equal to one),
$\boldsymbol{\mathcal E}$ is an {\it external electric field}
calculated in a comoving frame, $^L\mathbf E$ is a vector field,
called {\it local electric field}, accounting for the interaction
between the polarization of different species with the crystal
lattice, $\nabla\cdot^L \mathbb E$ is the divergence of a rank-two
tensor $^L\mathbb E$, called {\it local electric field tensor},
which accounts for the so-called {\it shell-shell interaction} and
finally $^L \mathbb E$ is responsible for the typical ferroelectric
ordering. The master equation (\ref{pi}) fixes the behavior of the
state variable $\boldsymbol{\pi}$ in terms of the electric fields
$^L\mathbf E$ and $^L\mathbb E$. Fields $^L\mathbf E$ and $^L\mathbb
E$ play the role of internal variables. This equation resembles
Newton equation law of motion.

 The internal energy
balance has  the form:

\beq\label{intenb} \rho\dot\epsilon=p_{(i)}-\nabla\cdot(\mathbf
q-\boldsymbol{\mathcal P}) \eeq where $\boldsymbol{\mathcal P}$ is
the Poynting vector and the work power of internal forces
$p_{(i)}$ is given  by

\beq\label{intpow} p_{(i)}=\boldsymbol\sigma : \mathbf
L-\rho{}^L\mathbf E : \dot{\boldsymbol\pi}+{}^L\boldsymbol{\mathbb
E} : \nabla\dot{\boldsymbol\pi}\eeq

By using the relation $\mathbf L=\dot{\mathbf F}\mathbf F^{-1}$ we
can rewrite (\ref{intpow}) as:

\beq\label{intpow1} p_{(i)}=\boldsymbol\sigma : {\mathbf F}^{-1}
:\dot{\mathbf F} -\rho{}^L\mathbf E
:\dot{\boldsymbol\pi}+{}^L\boldsymbol{\mathbb E} :
\nabla\dot{\boldsymbol\pi}.\eeq

Then, the dynamical system describing evolution of the state fields
of the system \cite{FRR} is the following:

\beq
\begin{cases}\label{dynamical}
 \dot{\mathbf F}= \mathbf L \mathbf{F}\\
 \dot\epsilon=\boldsymbol\sigma :{\mathbf
F}^{-1} : \dot{\mathbf F} -\rho{}^L\mathbf E :
\dot{\boldsymbol\pi}+{}^L\boldsymbol{\mathbb E} :
\nabla\dot{\boldsymbol\pi}-\rho^{-1}\nabla\cdot(\mathbf
q-\boldsymbol{\mathcal
P})  \\
  \dot{\mathbf H}= \mathbf q \\
  \dot{\boldsymbol\pi}= \mathbf u\\
  \dot{\mathbf u}= \boldsymbol{\mathcal
E}+{}^L\mathbf
E+\rho^{-1}(\nabla\cdot{}^L\boldsymbol{\mathbb E}) \\
  \dot{\nabla\boldsymbol\pi}= \nabla\mathbf u\\
  \dot{\nabla\mathbf u}= \nabla\cdot\mathbf J_{\nabla\mathbf
u}+\sigma_{\nabla\mathbf u},
\end{cases}\eeq
\newline

where $\mathbf J_{\nabla\mathbf u}$ and $\sigma_{\nabla\mathbf u}$
are both phenomenological quantities representing, respectively, the
current and the source terms associated with $\nabla\mathbf u$. The
introduction of the variable $\mathbf u=\dot{\boldsymbol\pi}$ has
been used so to obtain a first order dynamical system.
\par

 We finally present the entropy balance (\ref{entrobal1}) in the form:

\beq \small\dot s-
{\theta}^{-1}\dot\epsilon+{\rho\theta}^{-1}\boldsymbol\sigma :
{\mathbf F}^{-1} : \dot{\mathbf F}-\theta^{-1}{}^L\mathbf E :
\dot{\boldsymbol\pi}+(\rho)^{-1}{}^L\boldsymbol{\mathbb E} :
\nabla\dot{\boldsymbol\pi}-\rho^{-1}\mathbf
q\cdot\nabla\theta^{-1}-\rho^{-1}\nabla\cdot(\theta^{-1}\boldsymbol{\mathcal
P}+\mathbf k)=\Xi \eeq

where the general relation (\ref{entrobal}) for the entropy flux has
been assumed.

Then, the infinitesimal entropy production along a process is
given by the integral of the following 1-form:

$$ d\sigma=d s-
{\theta}^{-1}d\epsilon+{\rho\theta}^{-1}\boldsymbol\sigma :
{\mathbf F}^{-1}: d{\mathbf F}-\theta^{-1}{}^L\mathbf E:
d{\boldsymbol\pi}+ {\boldsymbol\pi}+$$
\beq\label{entrform}\quad\quad+\rho^{-1}{}^L\boldsymbol{\mathbb E}
: \nabla d-\rho^{-1}\mathbf
q\cdot\nabla\theta^{-1}dt-\rho^{-1}\nabla\cdot(\theta^{-1}\boldsymbol{\mathcal
P}+\mathbf k)dt\eeq

In such a case the evolution of the material properties is described
by the following functions of time forming the state space \beq
B=(\epsilon,\mathbf F,\boldsymbol{\mathcal H};\boldsymbol\pi,
\nabla\boldsymbol\pi,
\dot{\boldsymbol\pi},\dot{\nabla\boldsymbol\pi};t).\eeq

The vector field $\boldsymbol{\mathcal H}$ has been introduced
above.

The right side of (\ref{entrform}) can be considered as the exterior
1-form:

\begin{multline}\eta=\theta^{-1} d\epsilon-{\rho\theta}^{-1}\boldsymbol\sigma:{\mathbf
F}^{-1}:d\mathbf F +\theta^{-1}{}^L\mathbf E:
d\boldsymbol\pi-{\rho\theta}^{-1}{}^L\boldsymbol{\mathbb
E}:d\nabla\boldsymbol\pi+\boldsymbol\beta\cdot d\boldsymbol{\mathcal H}+\\
-\rho^{-1}(\theta^{-1}\nabla\cdot\boldsymbol{\mathcal
P}+\nabla\cdot\mathbf k )dt,
\end{multline}

remembering the definition $\boldsymbol
\beta=-\rho^{-1}\nabla\theta^{-1}$.

 We notice that the
Poynting vector together with the $\mathbf k$ quantity determine
 outside contributions in the entropy production. These arguments lead
us to the introduction of the large space of the variables:

\beq \mathcal B=\Big[s\cup
B\cup\Big(\frac{1}{\theta},-{\rho\theta}^{-1}\boldsymbol\sigma\cdot{\mathbf
F}^{-1},\frac{1}{\theta}{}^L\mathbf
E,-\frac{1}{\theta}{}^L\boldsymbol{\mathbb
E},-\nabla\theta^{-1},-\nabla\cdot \mathbf
k,-\frac{1}{\theta}\nabla\cdot\boldsymbol{\mathcal P}\Big)\Big].
\eeq

The exterior 1-form $\omega=ds-\eta$ defines the contact structure
in the space $\mathcal B$. We notice that the differential of the
form $\omega$ is given by

\beq\Omega=d\omega=-d\eta.\eeq

If $\eta$ is closed then, locally,  $\eta=d  U$ for some functions
$ U\in C^\infty(B)$. Assuming that the state space $B$ is simply
connected the question of the existence of the entropy as function
of the state variables rely on the property of the form $\eta$ to
be closed, i.e. $d\eta=0$. This gives the reason to analyze the
conditions for closeness of this 1-form.

Closeness of the form gives the potential $ U(\epsilon,\mathbf
F,\boldsymbol\pi,\nabla\boldsymbol\pi,\boldsymbol{\mathcal H},t)$ on
$B$ so that

\beq
\begin{cases}
\theta^{-1}=\partial_\epsilon  U,\\
\\
-(\rho\theta)^{-1}\boldsymbol\sigma\cdot{\mathbf
F}^{-T}=\partial_F  U,\\
\\
^L\mathbf E=\theta\partial_\pi  U,\\
\\
-(\rho\theta)^{-1}{}^L\boldsymbol{\mathbb E}=\partial_{\nabla\pi}U,\\
\\
\boldsymbol\beta=\partial_\mathcal H U,\\
\\
\boldsymbol\beta\nabla\cdot \boldsymbol{\mathcal P}-\rho^{-1}
 \nabla\cdot\mathbf k=\partial_t\mathcal U.
\end{cases}\eeq

By using the identity $\theta^{-1}\nabla\cdot\mathbf{\mathcal
P}=\nabla\cdot(\theta^{-1}\boldsymbol{\mathcal
P})-boldsymbol{\mathcal P}\cdot\nabla\theta^{-1}$ the last equation
of the above system takes the form

\beq \boldsymbol{\mathcal P}\cdot
\boldsymbol\beta-\rho^{-1}\nabla\cdot(\theta^{-1}\mathbf{\mathcal
P}+\mathbf k)
      =\partial_t  U.
\eeq

Finally, if we assume the full integrability of $\eta$, we get the
system (\ref{dynamical}) in terms of potential $  U$ of the
entropy form as:

\beq
\begin{cases}\label{dynamical1}
 \dot{\mathbf F}= \mathbf L\mathbf F,\\

 \dot\epsilon= \theta\partial_F U:\mathbf L\mathbf F
-\rho\theta\partial_\pi  U\cdot\mathbf u
-\rho\theta\partial_{\nabla\pi}
 :\nabla\mathbf u
 -\rho^{-1}\nabla\cdot\mathbf
q-\theta(\rho\nabla\cdot\mathbf k+\partial_t  U),\\
  \dot{\boldsymbol{\mathcal H}}= k^{-1}\rho\partial_{\mathcal H}  U,\\
\dot{\boldsymbol\pi}= \mathbf u,\\
\dot{\mathbf u}= \boldsymbol{\mathcal E}+\theta\partial_\pi  U-\rho^{-1}(\nabla\cdot\theta \rho\partial_{\nabla \pi}  U), \\
  \dot{\nabla\boldsymbol\pi}= \nabla\mathbf u,\\
  \dot{\nabla\mathbf u}= \nabla\cdot\mathbf J_{\nabla\mathbf
u}+\sigma_{\nabla\mathbf u}.
\end{cases}\eeq

In order to close this system of equations in the space of basic fields $B$ one would have to use constitutive relations for the following entries in the system:
\begin{enumerate}
\item The terms $L,\nabla\cdot\mathbf q$ as in the simple model
\cite{CO}, \item the term $\nabla\cdot\mathbf k$ which would be
present if mixed dissipative processes would go in the system,
that one usually is determined using a dissipative potential,
\cite{M1}, \item gradients of variables appeared in
$\nabla\cdot\theta \rho\partial_{\nabla \pi}\mathcal U),$ \item
flux and production terms in the last equation.
\end{enumerate}

It is hardly possible to do this in some regular and relatively
simple way.  This problem is similar to the problem of closeness of
systems of equations for momenta in Statistical Mechanics or
Rational Extended Thermodynamics.\par It seems more reasonable to
lift the problem to the larger space,i.e. the extended
thermodynamical phase space $\mathcal{B}$ where several models of
thermodynamical processes were suggested, \cite{EMS, Ha, Ha2}, etc.
\par To prepare the framework for such development, we present, in
the next part of this work, the geometrical scheme introducing the
entropy form, its integrability (closeness) into the conventional
thermodynamical phase space, extended, whenever necessary, by the
time $t$ added to the list of extensive variables.

\vskip1cm

\addtocontents{toc}{\textbf{Part II. Contact geometry of entropy form.}}
\centerline{\textbf{Part II. Contact geometry of entropy form.}}

\section{Contact structure of homogeneous thermodynamics.}

In this section we briefly recall the \textbf{standard contact structure} of
homogeneous thermodynamics in the thermodynamical phase space
introduced by C.Caratheodory and developed by R.Hermmann and R. Mrugala
(\cite{Ca,H}).

A phase space of the \emph{homogeneous thermodynamics} (\textbf{thermodynamical
phase space}, or TPS) is the (2n+1)-dimensional vector space
$\mathcal{P}=\mathbb{R}^{2n+1}$ endowed with the \textbf{standard contact structure}
(\cite{AG,H}). Contact structure is defined by the (contact) 1-form $\vartheta$ such that
exterior product of $\vartheta$ and $n$ copies of its differential $d\vartheta$ is nonzero $(2n+1)$-form:
\[
\vartheta \wedge d\vartheta \wedge\ldots \wedge d\vartheta \ne 0
\]
By D'Arbois Theorem, \cite{A}, there is a choice of coordinates (local in a general manifold and global for the standard contact structure)
 $(z; (q^1 ,p_1),\ldots (q^n ,p_n ))$ such that the $\vartheta$ takes the form
\begin{equation} \vartheta
=dz-\sum_{i=1}^{n}p_{i}dq^{i}.
\end{equation}

The horizontal distribution $D=Ker(\vartheta)$ of this structure is generated by two
families of vector fields
\[
D=<\partial_{p_{l}},\ \partial_{q^{i}}+p_{i}\partial_{s}>. \]

The 2-form
\[
\Omega =d\vartheta =-\sum_{i=1}^{n}dp_{i}\wedge dq^{i}
\]
is a nondegenerate, symplectic form on the distribution $D$.\par

The Reeb vector field, uniquely defined as the generator $\zeta $ of
the 1-dim \emph{characteristic distribution} $ker(d\vartheta)$
satisfying $\vartheta(\zeta )=1$, is simply
\[
\zeta = \partial_{s}.
\]

\section{Gibbs space. Legendre surfaces of equilibrium.}
Concrete thermodynamical systems are determined by their \textbf{constitutive relation}, which, in their
conventional form determine the value of a \textbf{thermodynamical potential} $z=E(q^{i})$ as the
function of n (extensive) variables $q^{i}$ of a D'Arbois canonical coordinate system $(z; (q^1 ,p_1),\ldots (q^n ,p_n ))$. Dual, intensive, variables are determined then as the partial derivatives of the thermodynamical potential by the extensive variables: $p_{i}=\frac{\partial E}{\partial
q^{i}}$.\par
  Geometrically, a constitutive relation is determined as a {\bf Legendre submanifold}
(maximal integral submanifold) $\Sigma_{E}$ of the contact form
$\vartheta$.  Locally a Legendre submanifold $\Sigma$ is determined by a choice of canonical coordinates
$(z; (q^1 ,p_1),\ldots (q^n ,p_n ))$ such that an open subset $U\subset \Sigma$ projects diffeomorphically to
the space $X$ of variables $q^{i}$. In terms of these coordinates $\Sigma$ is defined  in the open domain $U$ as follows:
\beq
\Sigma =\{(z,q,p)\in P\vert z=E(q^i ),p_{i}=\frac{\partial E}{\partial q^i}\}.
\eeq
more about local presentation of Legendre submanifolds and their properties see \cite{A,AG}.\par
Space $G$ of variables $z,q^{i},\ i=1,\ldots ,n$ is, sometimes, named the {\bf Gibbs
space (bundle)} of the thermodynamical potential $E(q^{i})$.
Thermodynamical phase space $(\mathcal{P},\vartheta )$ (or, more
precisely, its open subset) identifies with the {\bf first jet
space} $J^{1}(Y\rightarrow X)$ of the (trivial) line bundle
$\pi:G\rightarrow X$. Projection of $\Sigma_{E}$ to the Gibbs space
$Y$ is the {\bf graph} $\Gamma_{E}$ of the fundamental constitutive
law $E=E(q^{i})$.
\par

Another choice of the thermodynamical potential together with the
n-tuple of extensive variables leads to another representation of an
open subset of TPS $\mathcal P$ as the 1-jet bundle of the
corresponding Gibbs space. \par

The most commonly used thermodynamical potentials are: internal energy, entropy, free energy of Helmholtz,
enthalpy and the free Gibbs energy.\par

On the intersection of the domains of these representations, corresponding points are related by the {\bf
contact transformations} (see \cite{A,C}).\par

\begin{example}
As an example of such a thermodynamical system, consider the van der
Waals gas - a system with two thermodynamical degrees of freedom.
Space $\mathcal P$ is 5-dimensional (for 1 mole of gas) with the
canonical variables $(U,(T,S),(-p,V))$ (internal energy,temperature,
entropy, -pressure, volume), the contact form
\[
\vartheta =dU-TdS+pdV,
\]
and the fundamental constitutive law
\[
U(S,V)=(V-b)^{\frac{R}{C_{V}}}e^{\frac{S}{c_{V}}}-\frac{a}{V},
\]
where $R$ is the Ridberg constant, $c_{V}$ is the heat capacity at constant volume, $a,b$ are parameters
of the gas reflecting the interaction between molecules and the part of volume occupied by molecules
respectively, see \cite{C}.
\end{example}

\section{Extended thermodynamical phase space, its contact structure.}

Introduce the \textbf{extended thermodynamical phase space}
$\mathcal{P}^{2m+1}$ (ETPS).  This space contains $m$ physical
fields $q^i , i=1,\ldots ,m$ that may include, together with basic
fields (temperature, density, polarization vector, etc.) also their
space gradients. As an element new in comparison to the usual
thermodynamical phase space in $\mathcal P$, \textbf{time t may be included
 as the m-st variable $q^m =t$}. In addition, $m$ variables
$p_i$ dual to $q^i$ (including, possibly, $p_m$ dual to the time variable $q^m=t$)
and the thermodynamical potential (entropy in our work) $s$ are
considered as the variables in the space $\mathcal{P}$.  It is
convenient to consider both types of situations - where time is
included as an independent variable and where it is not (see Section 2 where the field $\boldsymbol{\mathcal H}$
 was introduced).\par
\begin{remark} We change notation of the coordinate $z$ of the canonical coordinate system into $s$  because in the case of MT-model, we
will consider entropy $s$ as the thermodynamical potential.
\end{remark}

ETPS $\mathcal{P}$ will be endowed with the standard contact structure

\beq \vartheta =ds-\sum_{i=1}^{m }p_i dq^i \eeq

As above we denote by $D=Ker (\vartheta )$ the horizontal distribution of the
contact structure $(\mathcal{P},\vartheta )$ and by $\xi$ the
corresponding Reeb vector field.

\begin{definition}
\begin{enumerate}
\item An \textbf{extended constitutive surface} (ECS) is the m-dim submanifold $\Sigma \subset \mathcal{P}$ such that the
restriction of the contact form $\vartheta$ to $\Sigma$ \textbf{is exact}:

\beq \label{ECS}\vartheta\vert_{\Sigma} =d(s-U) \eeq for a function
$U\in C^{\infty}(\Sigma )$.
\item If $\Sigma\subset \mathcal P$ is an extended constitutive submanifold and $U$ - function in the definition above, the function
\[
\sigma =s-U,
\]
defined on the surface $\Sigma$ is called the \textbf{entropy production potential}.
\end{enumerate}
\end{definition}

\begin{remark} It is convenient to write the function in the right side of (\ref{ECS}) as $s-U$
because this $U$ coincide with the potential of the entropy form
$\eta$ from the Part I.
\end{remark}

\begin{proposition}
For an arbitrary extended constitutive surface $\Sigma$ there exists
a Legendre submanifold $\Sigma_{0}$ of the contact structure
$\vartheta$ and a function $\sigma \in C^{\infty}(\Sigma_{0})$ such that the ECS $\Sigma$ is obtained from Legendre submanifold $\Sigma_{0}$
 by the shift by the flow of the Reeb vector field $\zeta$:
A point $(s,q^i,p_i)\in \Sigma_{0}$  is shifted with the
value of parameter equal to ($-\sigma(s,p_i,q^i)$:

\beq \Sigma =\{ exp(-\sigma(s,p,q)\zeta)(m),\ m=(s,p,q)\in \Sigma_{0}.\} \eeq
\end{proposition}

\begin{proof} If $(\Sigma, U)$ are respectively an ECS and the corresponding potential, extend the function $U$
out of the surface $\Sigma$ to some smooth function in
$\mathcal{P}$. The form $-\sum_{i=0}^{m }p_i dq^i $ in $\mathcal{P}$
vanishes on the submanifold $\Sigma$.\par

Consider the mapping $\phi: (s,p_i,q^i)\rightarrow
(s+\sigma,p_i,q^i)$ in a neighborhood of surface $\Sigma$, Let
$\phi(\Sigma)$ be the image of surface $\Sigma$ under this mapping.
Notice that the pullback of contact form $\vartheta$ under the
mapping $\phi$ is:
\[
\phi^{*}\vartheta =d(s-U)-\sum_{i=0}^{m }p_i dq^i .
\]
Let $\xi\in T_{x}(\Sigma)$ be a tangent vector to the surface $\Sigma$ at a point $x$.  Then
\[
0=<d(s-U)-\sum_{i=0}^{m }p_i dq^i,\xi >=<\phi^{*}\vartheta
,\xi>=<\vartheta ,\phi_{*x}\xi >.
\]

Therefore, surface $\Sigma_{0}=\phi(\Sigma)$ is Legendre surface of
contact structure $(\mathcal{P},\vartheta)$.  Therefore, the ECS
surface $\Sigma$ is obtained by the deformation of a Legendre
submanifold $\Sigma_{0}$ by the flow mapping $exp(-\sigma\zeta)$ of
the Reeb vector field $\zeta$ at the value of parameter $-\sigma$.
\end{proof}

\begin{remark} Function $U$ is defined up to addition of a constant. Another choice of $U$ leads to the shift of the
ECS by this constant in the direction of variable $s$, i.e. along
the trajectory of the Reeb vector field $\zeta$. Correspondingly,
the Legendre submanifold $\Sigma_{0}$ is shifted.  As a result, we
came up with the $(m+1)-dimensional$ submanifold $\Delta$ foliated
by the shifts of Legendre submanifold $\Sigma_{0}$ ECS $\Sigma_{U}$
and, transversally, by the phase curves of the Reeb vector field.
\end{remark}

\begin{corollary} With any admissible process $\chi: T\rightarrow \Sigma$ (see below the definition) there is related an
uniquely defined reversible process $\chi_{0}:T\rightarrow \Sigma_{0}$ defined by the condition:
\beq
\chi_{0}(t)=exp(\sigma(\chi (t))\zeta)\circ \chi(t),
\eeq
where
\beq
\sigma =s-U
\eeq
is the\textbf{ entropy production potential}.
\end{corollary}

Denote by $\pi_{\Sigma}:\Sigma \rightarrow X$ the projection of ECS to the base $X$ and by $\pi_{\Sigma_0}$ -
corresponding projection for associated equilibrium surface $\Sigma_{0}$.\par

If this projection is invertible on some open subset $W \subset \Sigma$, denote by $j_{\Sigma}:U\rightarrow W\subset \Sigma $
corresponding inverse mapping.  We will call mapping $j_{\Sigma}$ - the \textbf{characteristic embedding of} $\Sigma$.\par

Using the embedding $j$ one can directly relate the closeness of entropy form $\eta$ studied in Part I with the
requirement of integrability  of the contact form $\theta$ along the admissible dynamical processes $\chi (t)$:

\begin{proposition} Entropy form $\eta$ is closed on the base space $X$ if and only if
\beq \eta =j^{*}\vartheta\ (=dU) \eeq where $j:X\rightarrow
\mathcal{P}$ has the \emph{extended constitutive surface} as its
image.
\end{proposition}

A conventional way to specify such an extended constitutive surface
is to identify the ETPS $(\mathcal{P},\vartheta )$ with the 1-jet
space of the Gibbs bundle - a line bundle $G\rightarrow X$ over the
space $X$ of variables $(q^i ,t)$,see \cite{H}, (or simply $q^i$ in
the time-independent representation):

\beq
\begin{CD}
\mathcal P\simeq J^{1}{\pi}\\
@V\pi_{10}VV\\
G=R_{s}\times X \\
@V\pi VV \\
X
\end{CD}
\eeq

A choice of a section of the bundle $\pi$ - an entropy function $s=S(q^i ,t)$ allows to form the
Legendre submanifold $\Sigma_S = j^{1}(S)(X)$ - image of the space $X$ under the 1-jet section $j^{1}(S)$
of the bundle $\mathcal{P}\rightarrow G\rightarrow X$. Legendre submanifold $\Sigma_S$ constructed in such a way
projects diffeomorphically to the space $X$.\par

Choose next a function $\sigma_{0}\in C^{\infty}(X)$. Lift this function to the surface $\Sigma_{S}$ to get the function
\beq
\sigma =\sigma_{S}\circ \pi_{1}\vert_{\Sigma_S}.
\eeq

Now we \textbf{define the extended constitutive surface}

\beq
\Sigma_{S,\sigma}=\{z^* = exp(-\sigma(z))z \vert z\in \Sigma_{S}\}.
\eeq
\par

In this case the mapping $j(q)=exp(-\sigma(q))\circ j^{1}(S)(q)$ is
the characteristic embedding for ECS $\Sigma_{S,\sigma}$.\par

It follows from the local description of Legendre submanifolds of
contact structure (\cite{A,AG} that the following statement is valid

\begin{proposition}  Any extended constitutive submanifold $\Sigma \subset \mathcal{P}$ with the set of basic variables $q^i$ \textbf{locally}
has the form $\Sigma_{S,\sigma}$ for two functions $S,\sigma \in
C^{\infty}(X)$.
\end{proposition}

\begin{remark} For a general Legendre manifold, where the function ("potential") defining the variable of $z$ depends on $q^i, \ i\in I;p_j, j\in J$ for some decomposition $[1,m]=I\cup J$ of the set of indices from 1 to m, one can modify this definition accordingly to define corresponded shifted Legendre submanifold.
\end{remark}

Notice, that since a general Legendre submanifold of the contact manifold $(\mathcal{P},\theta )$ might have singularities or be given by multivalued function $S$, a general extended constitutive surface $\Sigma$ and its projection to $X$ may have singularities (Legendre singularities, see (\cite{AG}). A standard example of this kind is the projection of the constitutive surface of the van
der Waals gas to the pT plane, see \cite{C}.\par

Let $\chi:T\rightarrow X$ be a curve defining the process (evolution
of state). Combining it with the constitutive mapping $j_{S,\sigma}$ of a ECS $\Sigma_{S,\sigma}$ we get the
curve $\widehat\chi$ in the space $\mathcal P$.

\begin{definition}
A \textbf{thermodynamically (TD) admissible process} (with the entropy function $S$ and the entropy production potential $\sigma$) is the curve $\chi:T\rightarrow X$ such that for its lift $\widehat{\chi}:T\rightarrow \Sigma_{S,\sigma}$ given by
\[
\widehat{\chi}(t)=j_{S,\sigma}( \chi (t)),
\]
one has \beq \langle \vartheta
(\widehat{\chi}(t)),\widehat{\chi}'(t)\rangle = d\widehat{\sigma}
(\widehat{\chi}'(t))=\langle d\sigma(\chi (t)),\chi'(t)\rangle \geqq
0,\ \text{for\ all} \ t. \eeq

\end{definition}

Thus, the calculation of the entropy production in this model can be
done directly on the space $X$ of basic fields.  Change of entropy
along the process $\chi$ during the time interval $(t_{0},t_{1})$ is
equal to \beq \Delta s\vert_{t_0}^{t_1} =\Delta U+\Delta \sigma
=(U(t_1 )-U(t_0))+(\sigma (t_1 )-\sigma (t_0 )). \eeq

\vskip 1cm \textbf{R\`{e}sum\`{e}:} In a case of the closed  entropy
form $\eta$, entropy form model is (at least locally) presented by
two potentials - $(U,\sigma)$ or $(S=U+\sigma ,\sigma )$ where $U$
is the potential of the form $\eta$, $S$ is the entropy density and
$\sigma $ is the entropy production potential. Potential $U$ (or
$S$) is defined up to addition of a constant.\par Dynamical
evolution is presented by a thermodynamically admissible curves in
the space $X$ of basic fields or by a the TD-admissible curve on the
extended constitutive surface $\Sigma _{S,\sigma}$ in the extended
thermodynamical phase space $(\mathcal{P},\vartheta)$.

\par

\section{Entropy form as a flat connection in the Gibbs bundle.}

 Gibbs space $G$ is endowed with (local) coordinates $(s; q^i,i=0,\ldots, m )$ and
the corresponding frame $\partial_t ,\partial_{q^i},\partial_s $
together with the corresponding coframe $dq^i ,ds$.\par

Having the constitutional surface $\Sigma_{S,\sigma}$ and the
corresponding Legendre submanifold $\Sigma_{S}$ available we may
extend $\Sigma_{S}$ to the (m+1)-dim submanifold $\Lambda_{S}$,
possibly with singularities (or locally, without singularities) by
applying to the points of $\Sigma_{S}$ the flow of the Reeb vector
filed $\zeta$:
\beq \Lambda_{S}=\{exp((s-S(q))\zeta )j_{S}(q)\vert
(s,q)\in G\}. \eeq

 This definition of  $\Lambda_{S}$ allows to
define the smooth mapping

\[\lambda:G\rightarrow \Lambda_{S}:\ \lambda (s,q)= exp((s-S(q))\zeta )j_{S}(q),\]
which, away from the possible singularities of Lagrange submanifold
$\Sigma_{S}$ is correctly defined.

In this situation we can associate with the contact 1-form $\theta =ds-\sum_{i}p_i dq^i$ in the ETPS $\mathcal{P}$ the 1-form
\beq
\omega =\lambda^{*}\theta =ds-\eta =ds-\sum_{i=0}^{m}p_i(q) dq^i.
\eeq

Constructed 1-form defines the projector

\beq \Pi_v (\xi ) =(ds-\vartheta )(\xi )\partial_s ,\ \xi \in
T(G)\eeq

to the vertical subbundle $V(G)\subset T(G)$ of the tangent bundle of space $G$: $\Pi_v (\partial_s )=
\partial_s $.

This projector defines the \textbf{connection $\omega$ on the Gibbs line bundle}.\par

\emph{Horizontal space} of this connection is defined by the
condition $\Pi_v (\xi)=0 $. It consists of the tangent vectors $ \xi
=\xi^0 \partial_t +\xi^i \partial_{q^i}+\xi^s
\partial_s$ for which

\[ \xi^s =  {\hat p }_0 \xi^0 + {\hat p }_i \xi^i \Rightarrow  \xi =\xi^0 \partial_t +\xi^i \partial_{q^i}+({\hat p }_0 \xi^0 + {\hat p }_i \xi^i)
\partial_s =\xi^0 (\partial_t +{\hat p }_0 \partial_{ s })+\xi^i (\partial_{q^i}+{\hat p }_i \partial_{s}) ,\]
so, that

\beq\label{Hor} Hor(\omega) =\{<\partial_t +{\hat p }_0 \partial_{ s
},\partial_{q^i}+{\hat p }_i \partial_{s}>\}. \eeq

Coefficients of the projector form do not depend on $s$, thus the
connection defined in this way is the linear connection.  We will
call it $\eta_{\Sigma }$ since it is defined by the constitutive
relation $\Sigma$.

Curvature of the connection $\eta_{\Sigma }$ is zero

\beq \Omega =D_{\eta }\eta_\Sigma =-d\eta_\Sigma (\Pi_h \cdot ,\Pi_h
\cdot))=0\eeq since $\eta_\Sigma =dS$.

\subsection{General connection in Gibbs bundle as an entropy form.}

Consider a general connection in the Gibbs line bundle for a
situation where no $t$ is present in the list of canonical variables
(or simply take $t=q^{m+1}$, i.e.  defined by the 1-form \beq \eta
=p_{i}(s,q)dq^i, \eeq where functions $p_{i}(s,q^j)$ are arbitrary.
Connection form defining projection to the vertical subbundle
$\langle \partial_{s}\rangle \subset T(G)$ is \beq \omega =ds-\eta.
\eeq

Horizontal vector fields have the same form (\ref{Hor}) as above but
with $\xi_{0}=0$.  The curvature of this connection is determined by
the curvature form $\Omega =D_{\omega}\omega =d\omega (P_{h}\cdot\
,P_{h}\cdot\ )$ \cite{KN} Since

\beq d\eta =p_{i,s}ds\wedge dq^i +p_{i,j}dq^j\wedge dq^i, \eeq

 then
on the couples of basic horizontal vector fields
$\partial_{q^i}+{\hat p }_i \partial_{s},$ \beq \Omega
(\partial_{q^i}+{\hat p }_i \partial_{s} ,\partial_{q^j}+{\hat p }_j
\partial_{s})=(p_{j,s}p_{i}-p_{i,s}p_{j})+(p_{j,q^i}-p_{i,q^j}).
\eeq In this general case integrability conditions for the form
$\eta$ are not fulfilled and entropy potential $U$ is not defined
even if the entropy form $\eta$ is independent on variable $s$.\par
\begin{remark}  It would be interesting to see, possibly, on examples, the meaning of curvature in the case where entropy form is only partly integrable.
\end{remark}

\section{Entropy form geometrically: conditions of integrability. }

In a number of works, see \cite{DFR1,DFR2,DFRe,FRR} and the literature cited thereon, the closeness conditions of
the entropy form were calculated for different thermodynamical
systems. Here we show that these conditions take simple abstract
form in terms of the contact structure of ETPS $\mathcal{P}$. In these form the integrability conditions may be
 useful for writing down and studying similar
conditions for specific physical systems.\par

Consider the 1-form in the extended base space with variables
$(q^i,t)$ \beq \eta' = p_{i}(q^i,t)dq^i-H(q^i,t)dt. \eeq

Let the form $\eta'$ be closed: $d\eta'=0$. Then
\[\small
0=(p_{i,q^j}dq^j +p_{i,0}dt)\wedge dq^i-(H_{,q^j}dq^j +H_{,t}dt )\wedge dt=
-(p_{i,q^j}-p_{j,q^i})dq^i \wedge dq^j -(p_{i,0}+H_{,q^i})dq^i\wedge dt.
\]
Thus, \textbf{closeness condition for the 1-form $\eta'$ is
equivalent to the fulfillment of the following relations:}

\beq
\begin{cases}
p_{i,q^j}-p_{j,q^i}=0,\ \forall i,j;\\
p_{i,t}+H_{,q^i}=0,\ \forall i.
\end{cases}
\eeq

and, \textbf{locally, is equivalent to the existence of the potential} $U$:
\beq
\begin{cases}
p_{i}=\frac{\partial U}{\partial q^i},\\
H=-\frac{\partial U}{\partial t}.
\end{cases}
\eeq where the function $U(q^i,t)$ is defined uniquely up to adding
of an arbitrary constant.\par

Let $s\rightarrow (q(s),p(s)t(s)$ be a parameterized curve in the phase space.  Then, we have the relation between rates of change of $p$ and $q$ variables:

\[
\partial_{s}{p}_{i}=U_{,q^i q^j}\partial_{s}{q}^j +U_{,q^i t}\partial_{s}t.
\]

In particular, for a curve $t\rightarrow (t,q(t),p(t))$  we have
\beq
\partial_{t}{p}_{i}=U_{,q^i q^j}\partial_{t}{q}^j +U_{,q^i t}.
\eeq
These relations are of geometrical nature and has to be compatible with any
 dynamical evolutional model of the material point.\par

\begin{remark}
Hessian matrix $R_{ij}=U_{,q^i q^j}$ of the entropy potential $U$ defines the \emph{thermodynamical metric},\cite{Mru} , coinciding with the Ruppeiner metric \cite{Ru} (defined by the Hessian of the entropy $S$) on the constitutive surface $p_i =\frac{\partial U}{\partial q^i}$ where the entropy production $\sigma$ is constant.  Equation (10.4) represents the relation between tangent and cotangent vectors to the process curve on the base $X$ defined by this metric.\par
Condition of non-degeneracy of this metric is:
\[
det(U_{,q^i q^j})=det (\frac{\partial p_{i}}{\partial q^j})\ne 0,
\]
i.e. represents local condition of invertibility of the godograph transformation $q^i \rightarrow p_i$.
\end{remark}

\section{Conclusion.}

In this work we analyzed integrability (closeness) conditions of the
entropy form, introduced by Coleman and Owen in the model of
material point.\par  Considering first the model of thermoelastic
point and then the thermoelastic dielectric material, we showed that
the entropy potential that is defined for the closed entropy form
determines some terms in the assumed dynamical system of
corresponding model and produces the constitutive relations for
temperature and its gradient.

\par On the other hand, dynamical systems for these systems stayed
unclosed and the arguments are presented that in the space of basic
fields it is impractical if not impossible to close this system.
\par That is why as an alternative to the basic space of fields, we
suggest to use extended thermodynamical phase space (ETPS)
$(\mathcal{P},\vartheta)$ with the canonical contact structure
$\vartheta$ similar to one studied in homogeneous thermodynamics.
\par In the second part of work we prove that the integrability
condition of the entropy form is equivalent to the conditions that
processes are confined to the "extended constitutive surface"
$\Sigma$ of ETPS. \par Structure of such surfaces is determined in
terms of Legendre submanifolds $\Sigma_{0}$ of the standard
contact structure of ETPS $\mathcal{P}$ and additional function
$\Sigma$ having sense of the entropy production potential. As a
result, a material point model in the ETPS with the integrable
entropy form is defined (at least locally) by two functions -
entropy $S$ - (function of basic fields, defining the Legendre
submanifold $\Sigma_{S}$) and the entropy production potential
$\sigma$ on such a submanifold defining the shift along the phase
curves the Reeb vector field that produces the extended
constitutive surface $\Sigma$. \par Thermodynamically admissible
processes are defined as the time parameterized curves on the
surface $\Sigma$ along which the value of entropy production
potential is increasing. This geometrical model of the material
point represents, in our opinion, an adequate framework for
construction of a thermodynamically admissible dynamical
model.\par

Next step in the development of a material point model(s)
would be to study the compatibility of known geometric models of irreversible thermodynamical processes -
Lagrangian systems with dissipative potential, \cite{M1}, metriplectic systems,
\cite{EMS} and the gradient relaxation processes of H. Haslash,\cite{Ha, Ha2} with the geometrical material point model.
 This problem will be studied elsewhere.

\end{document}